\documentclass[aps,prl,twocolumn,longbibliography,nobalancelastpage,floatfix]{revtex4-2}
\usepackage[T1]{fontenc}
\usepackage[utf8]{inputenc}
\pdfoutput=1 
\usepackage{graphicx}
\usepackage{amsmath}
\usepackage{amssymb}
\usepackage{amssymb}
\usepackage{bm}
\usepackage{physics}
\usepackage{comment}
\usepackage[colorlinks = true,
            allcolors = {Blue}]{hyperref}
\usepackage[capitalize]{cleveref}

\usepackage[dvipsnames]{xcolor}

\newcommand{\red}[1]{#1}

\let\Tr\relax\DeclareMathOperator\Tr{\mathrm{Tr}}%

\begin{document}

\author{Simon Kothe}
\affiliation{Department of Physics and SUPA, University of Strathclyde, Glasgow G4 0NG, United Kingdom}
\author{Christopher Oliver}
\affiliation{Department of Physics and SUPA, University of Strathclyde, Glasgow G4 0NG, United Kingdom}
\author{Peter Kirton}
\affiliation{Department of Physics and SUPA, University of Strathclyde, Glasgow G4 0NG, United Kingdom}
\date{\today}
	
\title{\texorpdfstring{$\mathcal{PT}$}--symmetry breaking phase transitions in an LMG dimer}

\begin{abstract}
The open Lipkin-Meshkov-Glick (LMG) model provides a prototype of a dissipative phase transition which can be analyzed using mean-field theory. By combining the physics of this model with those of a quantum analogue of a parity-time reversal symmetry breaking transition we analyze the steady state phase diagram of a pair of coupled LMG models. Their competition generates a complex phase diagram with multiple steady-state phases and emergent dynamical regimes absent from either constituent model, including chaos characterized by the maximum Lyapunov exponent. We show that the effects predicted from mean-field theory survive in the full quantum model. 
\end{abstract}

\maketitle

\paragraph{Introduction}

Open quantum systems which include the effects of both drive and dissipation can exhibit behavior that is much richer than that of their equilibrium counterparts. This is even more apparent in many-body systems~\cite{Fazio2024} where the interaction with the environment can produce novel dynamical behavior which is not possible in equilibrium, even stabilizing phases which cannot occur as ground states of equilibrium systems.

Typically, the dynamics of such systems are described by master equations, which allow a much richer phenomenology than is possible in equilibrium~\cite{Kessler2012, Minganti2018, Soriente2021, Damanet2024}. This is fundamentally due to the fact that the symmetries~\cite{Hannukainen2018, Altland2021, Kawabata2023, Sa2023, debecker_role_2025} possible are more complex than those that can arise when constrained by a Hamiltonian. The effects of dissipation on phases of matter have been explored in a variety of experimental platforms from cavity~\cite{Baumann2010, Cai2021} and circuit~\cite{Chen2023} QED to semiconductor microcavity polaritons~\cite{Rodriguez2017} and nonlinear optical cavities~\cite{Beaulieu2025}. \red{The ability to engineer steady states and critical points of driven-dissipative systems are of importance for quantum technologies, such as dissipative quantum computing~\cite{verstraete_quantum_2009, mirrahimi_dynamically_2014}, quantum sensing~\cite{ding_enhanced_2022, salvatori_quantum_2014} and quantum thermodynamics~\cite{ma_quantum_2017}.}
 
Much recent effort has been spent examining how non-equilibrium processes can alter phase transitions which are closely related to those found in quantum ground states. In the open Dicke~\cite{Garraway2011, Kirton2019} and dissipative LMG~\cite{Morrison2008, Morrison2008a, Ferreira2019, debecker_role_2025} models it is found that, while the critical points are shifted slightly by the dissipation, the states remain largely unchanged. There is, however, another class of transition where the physics is dominated by the drive and dissipation. The prototypical example of this kind of transition is that of the Scully-Lamb laser~\cite{Scully1997} where incoherent driving of the emitters leads to the emergence of a macroscopic photon field. In more exotic setups, this kind of physics can be explored in models with an absorbing state~\cite{Carollo2019, Gillman2019, Carollo2022}, models which break time-translation symmetry~\cite{Iemini2018} and models with the microscopic generalization~\cite{Prosen2012, Prosen2012a, Huber2020, Huber2020a, Nakanishi2022} of parity-time ($\mathcal{PT}$) symmetry~\cite{Ashida2020, Arkhipov2020}. 

\red{In this paper, we introduce a model which features both of these categories of phase transition and study the phase structure and dynamics which emerge from their interplay.} We begin with the dissipative version of the LMG model~\cite{Lipkin1965, Morrison2008, Morrison2008a, Ferreira2019} which has a transition between a paramagnetic and a ferromagnetic state as the nonlinearity in the Hamiltonian is varied. This is augmented with a $\mathcal{PT}$-symmetric partner giving a dissipation-dominated phase transition between a normal state with definite magnetization and an infinite-temperature state as the inter-spin coupling is increased. 

\paragraph{Model}

\begin{figure}
	\centering
	\includegraphics[width=0.5\textwidth]{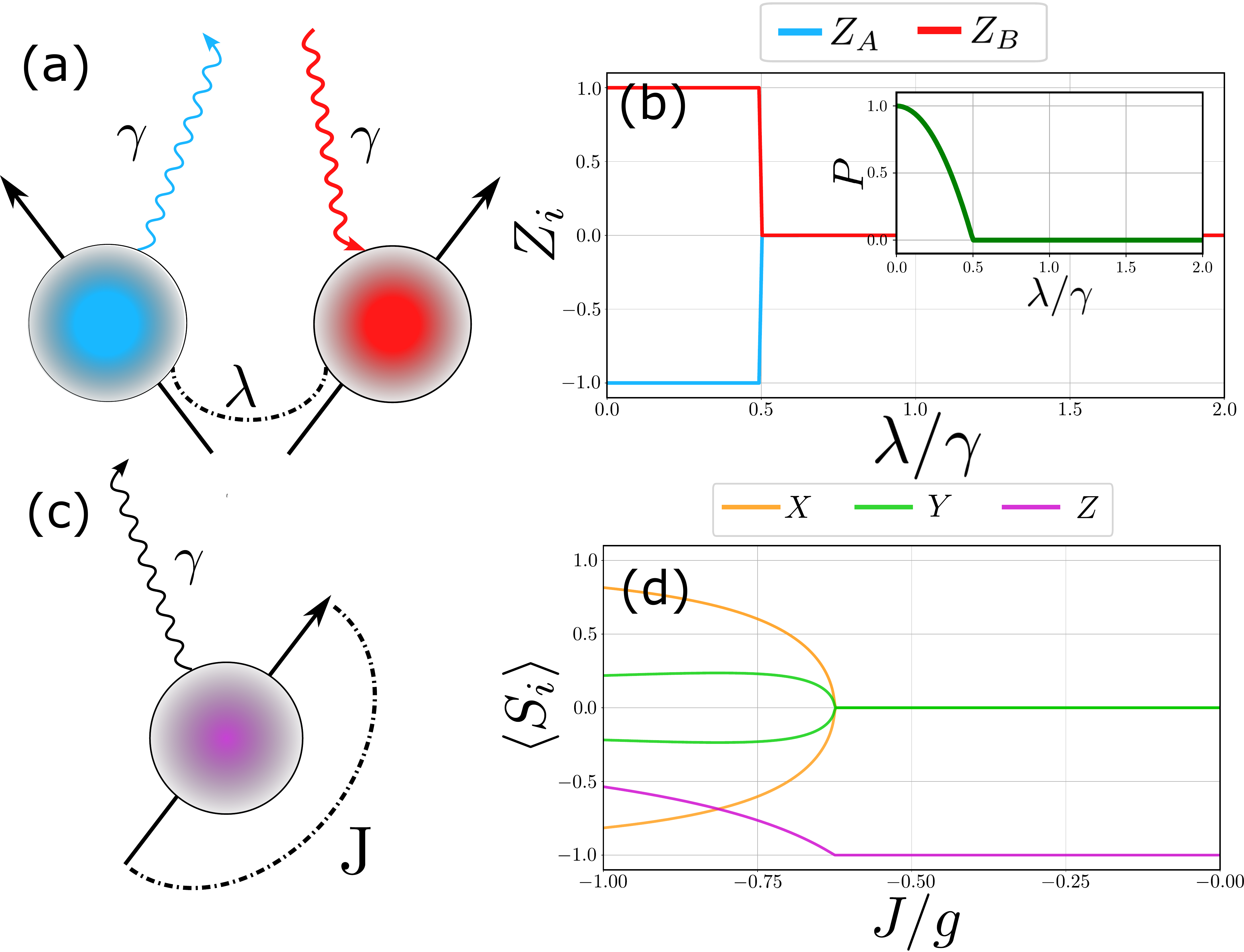}
	\caption{(a) Schematic diagram of the $\mathcal{PT}$ model. (b) Steady state magnetization and purity (inset) in the $\mathcal{PT}$ model as a function of the coupling $\lambda$. (c) Schematic diagram of the dissipative LMG model. (d) Steady state components of the spin magnetization in the dissipative LMG model at $\gamma=0.5g$. 
 \label{fig:model}}
\end{figure}

 We begin by briefly reviewing the phase transitions found in the $\mathcal{PT}$ and dissipative LMG models which underlie our main results. The $\mathcal{PT}$ model consists of two coupled spins, one of which absorbs excitations from the environment while the other emits excitations into it. The model is described by the master equation
\begin{equation}
    \dot \rho = -i\left[H_{PT},\rho\right] + \frac{\gamma}{S}\left(\mathcal{D}\left[S^-_A\right] + \mathcal{D}\left[S^+_B\right]\right),
\end{equation}
where $S_{A/B}^{+(-)}$ is the raising (lowering) operator for the collective spins, labeled $A$ and $B$, which are of length $S$.  The coherent term is $H_{{PT}} = \lambda(S^+_AS^-_B + S^-_AS^+_B)/2S$ and the Lindblad superoperators are given by
$
\mathcal{D}\left[O\right] = O\rho O^{\dagger} - \frac{1}{2}\lbrace O^{\dagger}O,\rho\rbrace .
$ 
This model is shown schematically in Fig.~\ref{fig:model}(a). It has a symmetry when swapping the labels of the sites (parity) and exchanging gain for loss by replacing $c\to c^\dagger$ in the jump operators.  Models with this symmetry generically have steady-state phase transitions where, on increasing $\lambda$, the state changes
from being dissipation dominated, with different magnetization on each site, to one where the coherent dynamics induce an infinite temperature maximally mixed state~\cite{Huber2020, Arkhipov2020, Nakanishi2022}. In the limit $S\to\infty$ this transition becomes discontinuous, corresponding to the semiclassical $\mathcal{PT}$ symmetry breaking transition of a non-Hermitian Hamiltonian.  This can be seen in Fig~\ref{fig:model}(b) where we show the magnetizations, $Z_A$ and $Z_B$ of the two spins as the coupling strength is varied.  In the inset to Fig.~\ref{fig:model}(b) we also show how the purity of the steady state changes. At $\lambda=0$, the system is in the pure state where spin-$A$ points up and spin-$B$ down, $\ket{\Downarrow \Uparrow}$ while above $\lambda=0.5\gamma$ the steady state is an equal mixture of all possible spin configurations~\cite{Huber2020, Huber2020a}. 

We also use the dissipative LMG model~\cite{Morrison2008, Morrison2008a, Ferreira2019}. It consists of a nonlinear spin coupled to a dissipative environment. The master equation describing this is
\begin{equation}
\dot \rho = -i\left[H_{LMG},\rho\right] + \frac{\gamma}{S}\mathcal{D}\left[S^-\right],
\end{equation}
where the Hamiltonian, $H_{LMG} = -{J}S_x^2/S - gS_z$, has competition between the nonlinearity, $J$, and magnetic field $g$ (Fig.~\ref{fig:model}(c)). 
This model has a $\mathbb{Z}_2$ symmetry corresponding to a rotation by $\pi$ in the $x$-$y$ plane. The ground state phase transition, which spontaneously breaks this symmetry, is between a paramagnet for small $|J|$, and a ferromagnet when $|J|$ is large. This carries over to the steady state of the dissipative model with a small correction to the location of the critical point, weakly dependent on the loss rate. Results for the individual components of the magnetization in the thermodynamic limit, $S\to\infty$, are shown in Fig.~\ref{fig:model}(d). For $J \approx 0$ the steady state has the spin pointing downwards, whereas at large negative $J$ two solutions for the in-plane magnetizations $\langle S_x\rangle$ and  $\langle S_y\rangle$ emerge. Note $\langle S_y\rangle$ is finite because of the presence of the dissipation which tilts the direction of the spin in the $x$-$y$ plane.  

The model which we focus on here builds on these two parts by combining a pair of LMG models, one dissipative, one driven, which are coupled via the $\mathcal{PT}$ Hamiltonian above. This full model retains the relevant features of both phase transitions. Our goal is to investigate the competition between them. The full Hamiltonian is
$
    H = H_{LMG}^A\otimes I^B + I^A\otimes H_{LMG}^B + H_{PT},
$
and the dynamics are described by the master equation
\begin{equation}
\dot\rho = -i\left[H,\rho\right] + \frac{\gamma}{S}\left(\mathcal{D}\left[S^-_A\right] + \mathcal{D}\left[S^+_B\right]\right)\, .
\end{equation}
When the nonlinearity, $J=0$, this reduces to the $\mathcal{PT}$ model described above and, when the coupling $\lambda=0$, this reduces to a pair of uncoupled LMG models.

\paragraph{Mean-Field Equations}

To analyze the behavior of this model we construct mean-field equations of motion. These are exact, for the underlying models, in the thermodynamic limit $S\to\infty$~\cite{Huber2020, Morrison2008}, and more generally for models in this class~\cite{Fiorelli2023, Wright2023} giving us confidence in their validity. For compactness we define e.g~$X_A = \Tr[S_x^A\rho]/S$, the normalized expectation values for the spin components such that $-1\leq X_A\leq 1$. The products which appear are broken using the mean-field ansatz $\Tr[S_x^AS_y^B\rho]/S^2 \simeq X_AY_B$. With this we obtain a set of 6 coupled non-linear equations, 3 for each site
\begin{align}
\dot{X}_A &= gY_A + \lambda Z_AY_B + \gamma X_AZ_A\, \label{eqn:MF1} ,\\
\dot{Y}_A &= -gX_A + 2J Z_AX_A - \lambda Z_AX_B + \gamma Y_AZ_A\, ,\\
\begin{split}
    \dot{Z}_A &= -2JY_AX_A - \lambda(X_AY_B - Y_AX_B) \\ &\hspace{12em} - \gamma(1 - Z_A^2)\, ,
\end{split}\\
\dot{X}_B &= gY_B + \lambda Z_BY_A - \gamma X_BZ_B\, ,\\
\dot{Y}_B &= -gX_B + 2J Z_BX_B - \lambda Z_BX_A - \gamma Y_BZ_B\, ,\\
\begin{split}
\dot{Z}_B &= -2JY_BX_B + \lambda(X_AY_B - Y_AX_B) \\ &\hspace{12em} + \gamma(1-Z_B^2)\, . \label{eqn:MFn}
\end{split}
\end{align}
Note that, because of our choice of scaling, the equations are all independent of the size of $S$. 
We first analyze the steady-state phase diagram of the model described by this set of equations.

The equations above always have a normal state solution with $X_A=X_B=Y_A=Y_B=0$ and $Z_A=-1$, $Z_B=1$. This corresponds to the dissipation dominated state of the $\mathcal{PT}$ model and the paramagnetic state of the individual LMG models. Therefore, this state is stable when both $\lambda\ll g$ and $|J|\ll g$. Away from this region, while the normal state is still a solution to the mean-field equations, it is no longer stable. Hence, to obtain the full phase diagram we must categorize all of the possible solutions to the mean-field equations and analyze their stability. While some analytic results are available, for example for the LMG model~\cite{Morrison2008}, in general this must be done numerically. \red{To find these solutions, we numerically search for steady-state solutions of Eqs.~\eqref{eqn:MF1}--\eqref{eqn:MFn} for each point in parameter space, check their stability from the eigenvalues of the Jacobian matrix and count the number of distinct solutions.}

\begin{figure}
    \hspace*{-0.5cm}   
	\includegraphics[width=\columnwidth, angle=90]{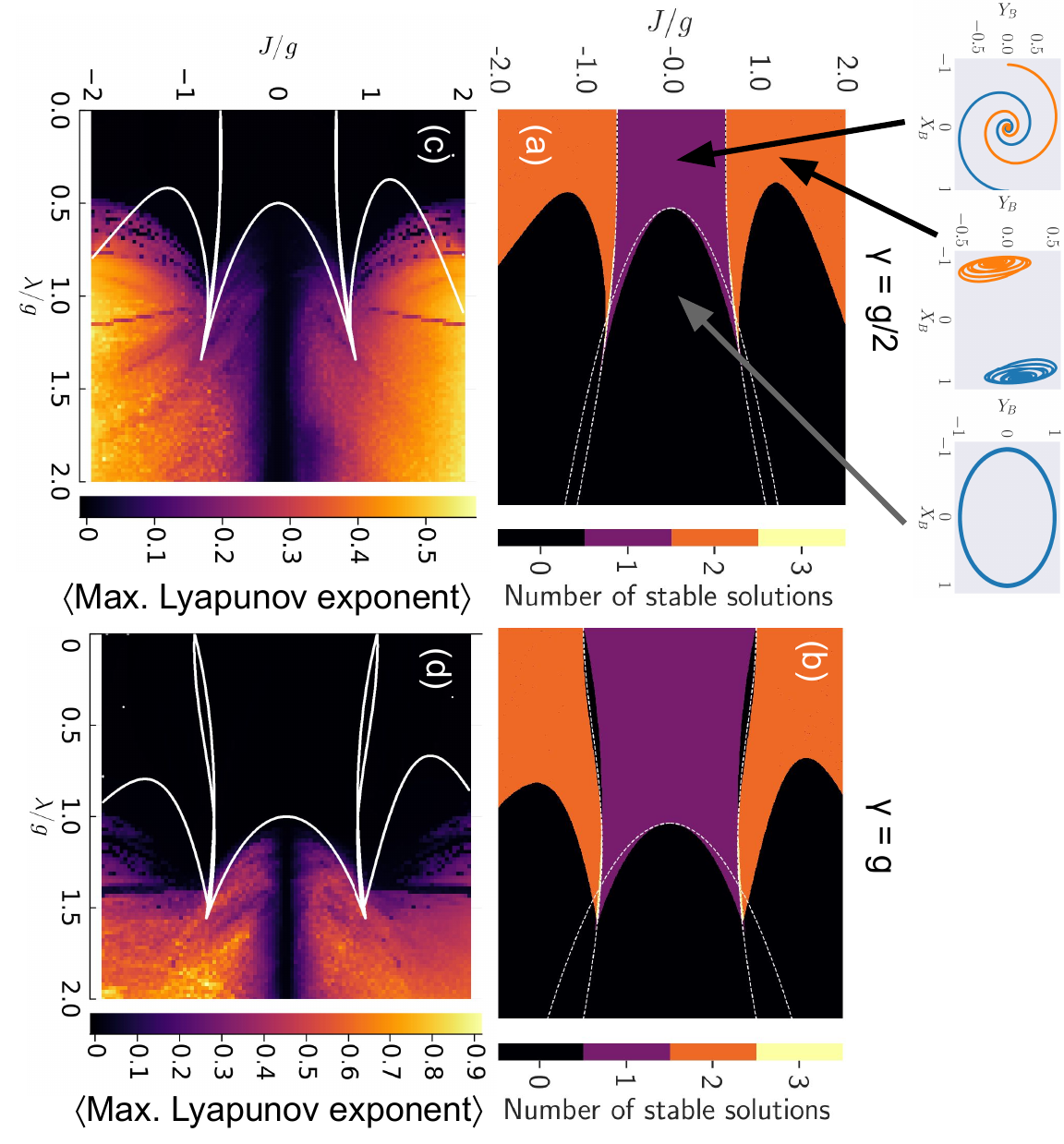}
	\caption{\red{Mean-field phase diagrams obtained from Eqs.~\eqref{eqn:MF1}--\eqref{eqn:MFn}. (a) and (b) show the number of stable fixed points for different values of $\gamma$ as a function of $J$ and $\lambda$, including sample dynamics for different phases. 
    Analytical approximations for the phase boundaries are shown as white dashed lines (Eqs.~\eqref{eqn:b1} and~\eqref{eqn:b2}). Panels (c) and (d) show the corresponding maximum Lyapunov exponent at each point in the phase plane. 
    Phase boundaries extracted from (a) and (b) are shown in white solid lines. 
    }}
 \label{fig:phase_diagram}
\end{figure}

The resulting phase diagram is shown in Fig.~\ref{fig:phase_diagram}(a) and (b) for $\gamma = g/2$ and $\gamma = g$ respectively. We show the number of stable solutions as a function of $J$ and $\lambda$. In the normal state only one fixed point is stable, this corresponds to the staggered magnetization phase described above. As $|J|$ is increased away from zero we reach phases where there are two stable solutions, related to the phase transition of the LMG model. When $J$ is positive spin-$B$ transitions into a ferromagnetic state, while when $J$ becomes negative spin-$A$ undergoes the same transition. This gives a pair of stable fixed points, one for each of the possible symmetry broken states.  

When the tunneling rate, $\lambda$, is large, we find that the mean-field equations have no stable fixed points and the long time behavior is given by a combination of limit cycles and more complex chaotic dynamics, discussed below. The existence of a limit cycle in this region is expected for the $\mathcal{PT}$-model. At large $\lambda$ the steady state of the full quantum model is given by the maximally mixed state. This state cannot be represented by mean-field theory since all mean-field states require $X_A^2+Y_A^2+Z_A^2=X_B^2+Y_B^2+Z_B^2=1$\footnote{Note that this is different to the condition imposed by the conservation of total angular momentum $\langle (S_A^x)^2\rangle + \langle (S_A^y)^2\rangle + \langle (S_A^z)^2\rangle = S(S+1)$ which is true for any state.}. In this regime mean-field theory predicts the existence of a limit cycle around the equator of the Bloch sphere which when averaged gives the result for the infinite temperature state. 

\red{We note two further features in the steady-state phase diagram on the boundaries between the normal and LMG phases. Firstly, particularly for large $\gamma$ (Fig.~\ref{fig:phase_diagram}(b)), a region with no fixed points emerges. The dynamics in these regions reveals behavior resembling critical slowing-down, where the $X$ and $Y$ components of the spin slowly approach, but do not reach, the fixed point expected in the normal phase. Secondly, parts of the boundary close to the $\mathcal{PT}$ phase show three stable fixed points. These correspond to the single fixed point in the normal phase and the pair of fixed points in the LMG phase, suggesting phase co-existence.} 

We find an analytic expression for the boundary to the LMG phase by locating where an eigenvalue of the Jacobian for the normal state becomes unstable, giving
\begin{equation}
{J_c^2} = \frac{(\gamma^2 -\lambda^2 +g^2)^2 + 4\lambda^2g^2}{4(\lambda^2 + g^2)},
\label{eqn:b1}
\end{equation}
i.e.~there are small corrections which, to lowest order, are quadratic in $\lambda$ to the location of these phase boundaries. 

In a similar way we can find the boundary of the limit cycle close to the line $J=0$. The resulting expression is more complex but to lowest order in $J$ we find
\red{\begin{equation}
    \frac{\lambda_c}{g}\simeq \frac{\gamma}{g}\left(1 + \frac{J^2}{2\gamma^2}\right).
    \label{eqn:b2}
\end{equation}}
These approximate results are shown as the white dashed lines in Fig.~\ref{fig:phase_diagram}(a) and (b). We see that they closely match the phase boundaries calculated numerically. 

The complexity of the mean-field phase diagram can be further explored by studying the dynamical behavior. In the normal state where only one solution is stable we find a single attractor for the dynamics which is reached from any set of initial conditions. In the two LMG phases there are a pair of attractors for the dynamics, corresponding to the two ferromagnetic states. This can be seen in the small panels in Fig.~\ref{fig:phase_diagram}, where the different colored lines show dynamics starting from different initial states. We also show the limit cycle at $J=0$ in the $\mathcal{PT}$ phase.

The behavior is much richer in the parts of the phase diagram where there are no fixed point solutions. As discussed above, exactly along the line where $J=0$ the $\mathcal{PT}$ phase transition is associated with the emergence of a regular limit cycle for all values of $\lambda$. The transition which occurs when increasing $\lambda$ at larger $|J|$ is much more complex. Here we find, at first, that each of the fixed points is associated with its own limit cycle which grows as $\lambda$ is increased. At some point these cycles merge and the dynamics fill phase space and are chaotic. 

\begin{figure}
	\centering
    \includegraphics[width=\columnwidth]{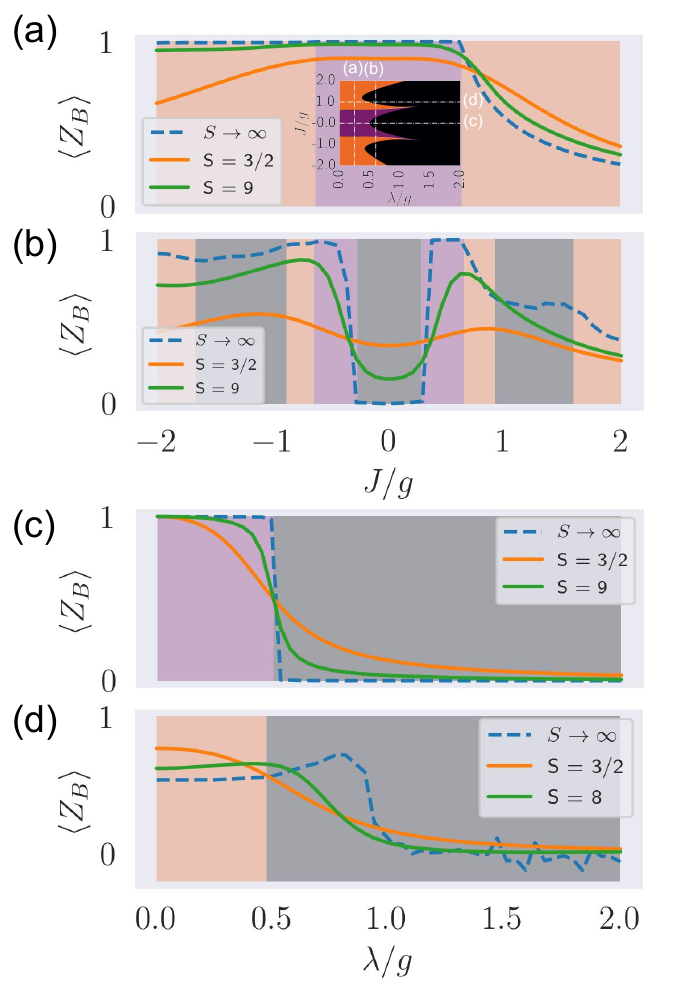}
	\caption{
    \red{Comparison of the results of full quantum simulations for the system sizes shown in the legend with the results of mean-field theory, for $\gamma = g/2$. The sweeps taken through the phase diagram are highlighted in the inset to panel c). In panels a) and b) we show sweeps at constant $\lambda$, while panels c) and d) show sweeps at constant $J$. All other parameters are the same as Fig.~\ref{fig:phase_diagram}}. 
 }
	\label{fig:exact}
\end{figure}

\red{We can shed further light on the richness of the dynamics in the $\mathcal{PT}$ phase by computing a dynamical phase diagram (Fig.~\ref{fig:phase_diagram}(c)--(d)). For each set of parameters, we numerically compute the maximum Lyapunov exponent~\cite{benettin_lyapunov_1980, geist_comparison_1990, datseris_dynamicalsystemsjl_2018, solanki_chaos_2025} and average it over symmetric initial conditions sampled from the equators of the Bloch spheres. 
}

\red{This procedure reveals rich structure in the $\mathcal{PT}$ phase. The Lyapunov exponent vanishes near the line $J = 0$, indicating the limit cycle dynamics. Curiously, near the boundary between the LMG and $\mathcal{PT}$ phases, some regions classified by the fixed-point analysis as stable nevertheless show a positive averaged Lyapunov exponent. The dynamics reveals different behavior depending upon the initial conditions. Some initial conditions go to the fixed points as expected, but others display complex oscillatory dynamics that does not settle down. This indicates additional complexity in the structure of the basins of attraction. Moreover, there are other structures that warrant exploration in future work, such as \textit{islands of stability} with reduced Lyapunov exponent. The complexity of the dynamics in the $\mathcal{PT}$ phase speaks to the power of our approach of systematically combining models with different kinds of phase transition, pointing the way towards a more general framework.}

\paragraph{Comparison to exact results}

To understand how the dynamical features emerge from a full quantum description we compare the results of the mean-field equations to those obtained from exact numerical diagonalization of the master equation. The size of the Hilbert space is that of two spin $S$ particles, $H_d=(2S+1)^2$, hence to find the steady state we have to work with a Liouville space of size $L_d=(2S+1)^4$. This means that only relatively small systems sizes are easily accessible. 

In Fig.~\ref{fig:exact} we take cuts through the phase diagram in Fig.~\ref{fig:phase_diagram}, comparing the value of $Z_B$ obtained from exact diagonalization at different values of $S$ with the mean-field results. \red{For parameter values without fixed points the mean-field results are obtained by averaging $Z_B$ over time and initial conditions}. The cuts in panels (a) and (c) of Fig.~\ref{fig:exact} are straightforward to interpret. Panel (a) shows the behavior as $J$ is varied at a fixed, small, value of $\lambda$. Here the physics is very similar to that of the uncoupled LMG model and we see the mean-field results predict that magnetization is fixed at 1 until a critical value of $J/g$ where it starts to decrease. The exact results agree with this picture with some finite-size corrections which reduce as $S$ is increased. Similarly, in Fig.~\ref{fig:exact}(c) where we show a sweep over $\lambda$ at fixed $J=0$. We see the magnetization changes sharply from 1 to 0 at $\lambda=g/2$, as expected for the $\mathcal{PT}$ model. In this case, the effect of finite $S$ is to smooth out the transition and to ensure that a unique steady state density operator is obtained, even when limit cycles are predicted for $S\to\infty$. 

In contrast, when the cut intersects with more phase boundaries the results are more complex. In Fig.~\ref{fig:exact}(b) we show the behavior at a fixed value of $\lambda$ which is large enough to intersect the regions without fixed points. We see that the quantum results match qualitatively with the mean-field predictions, but now the finite size effects are much more pronounced. The general trend, however, is still that the exact results approach those of mean-field theory as $S$ is increased. The noise in the mean-field results is due to imperfect averaging. Similarly, in Fig.~\ref{fig:exact}(d) we show a cut at fixed $J$ but now the value is large enough such that, at $\lambda=0$, the system is in the symmetry broken phase of the LMG model. The exact numerical results are in general agreement with the mean-field calculations, but again, the noise in the averaging procedure makes it difficult to be more quantitative than this. 

\begin{figure}
	\centering
	\includegraphics[width=0.85\columnwidth,scale=0.1]{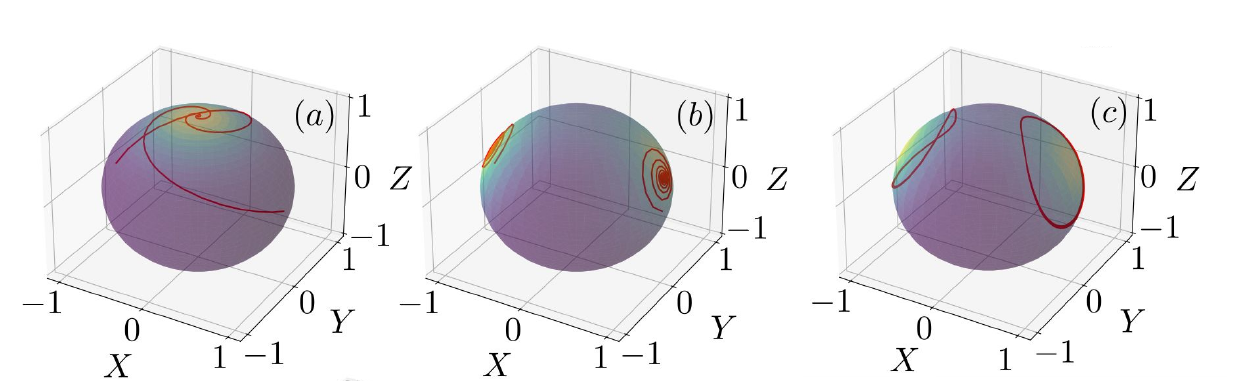}
	\caption{Numerically obtained spin Wigner function for three different parameter regimes with $S=3$. a) Normal phase at $J/g=\lambda/g =0$, (b) LMG phase at $J/g= 1.5$, $\lambda/g =0$ and (c) $\mathcal{PT}$ phase at $J/g= 1.5$, $\lambda/g =0.5$.}
	\label{fig:wigner}
\end{figure}

Finally, we note that it is possible to compute an analogue of the Wigner function for spins~\cite{Tilma2016}. This allows us to compare a phase space representation of the quantum state with the trajectories seen in mean-field theory. Examples of this are shown in Fig.~\ref{fig:wigner}. 
In panel (a) we show the spin Wigner function for $J=0$, $\lambda=0$, in this case the steady state points directly upwards towards the North pole of the Bloch sphere. Panel (b) shows a set of parameters in the orange region of Fig.~\ref{fig:phase_diagram}(a) where the mean-field prediction is that there are two steady state fixed points. We see this corresponds to two regions of high probability, at the locations predicted by mean-field theory.  In panel (c) we show a case from the region where the mean-field equations predicts a pair of limit cycles. We observe a more diffuse distribution of probability in the area where the limit cycles are present. We anticipate that this would be more apparent for larger systems.  

\paragraph{Conclusions}

\red{We have shown that coupling systems with distinct dissipative symmetry-breaking mechanisms can generate emergent dynamical phases absent from either constituent model. We demonstrated this using a pair of coupled nonlinear spins exhibiting both $\mathcal{PT}$ symmetry breaking as the coupling between the spins is changed and a conventional second-order phase transition driven by the nonlinearity. The resulting mean-field phase diagram contains phases inherited from the individual models, but the multiple limit cycles and chaotic regions, characterized by the maximum Lyapunov exponent, are not present in either parent model and emerge only from their competition. The main features are also present in finite-size exact quantum simulations.}

\red{Our work suggests a general framework for designing phases of driven-dissipative systems: copies of models with known phases can be coupled through interactions that themselves support distinct non-equilibrium transitions. This approach can be readily generalized, for example to lattice models where the phase transitions are richer, or to systems with more structured drive and dissipation, including non-Markovian baths. Determining the universality and finite-size scaling of the dynamical transitions found here is another important direction, likely requiring more specialized numerical approaches.}

\red{We anticipate that our model should be accessible using current technology. LMG-type models have been proposed or demonstrated in several experimental platforms, including superconducting qudits~\cite{champion_analog_2025}, ultracold Dysprosium~\cite{makhalov_probing_2019}, cavity QED~\cite{morrison_dynamical_2008}, circuit QED~\cite{xu_probing_2020} and an indirect realization using variational algorithms in a neutral atom tweezer array~\cite{chinnarasu_variational_2025}. Combined with developments in engineered dissipation across many platforms~\cite{harrington_engineered_2022}, this model should be feasible to realize experimentally. Such techniques may find application in quantum technologies where engineered steady states or critical points are resources, including dissipative quantum computing, quantum sensing and quantum thermodynamics~\cite{verstraete_quantum_2009, mirrahimi_dynamically_2014, ding_enhanced_2022, salvatori_quantum_2014, ma_quantum_2017}.}

\begin{acknowledgments}
    We acknowledge useful comments on a previous version of this paper from Fran\c{c}ois Damanet, Baptiste Debecker and Pablo Poggi. We thank Parvinder Solanki for fruitful discussions. SK acknowledges financial support from EPSRC (EP/ 517938/1). CO and PK acknowledge support from EPSRC (Grant No. EP/Z533713/1).
\end{acknowledgments}


\begin{thebibliography}{40}%
\makeatletter
\providecommand \@ifxundefined [1]{%
 \@ifx{#1\undefined}
}%
\providecommand \@ifnum [1]{%
 \ifnum #1\expandafter \@firstoftwo
 \else \expandafter \@secondoftwo
 \fi
}%
\providecommand \@ifx [1]{%
 \ifx #1\expandafter \@firstoftwo
 \else \expandafter \@secondoftwo
 \fi
}%
\providecommand \natexlab [1]{#1}%
\providecommand \enquote  [1]{``#1''}%
\providecommand \bibnamefont  [1]{#1}%
\providecommand \bibfnamefont [1]{#1}%
\providecommand \citenamefont [1]{#1}%
\providecommand \href@noop [0]{\@secondoftwo}%
\providecommand \href [0]{\begingroup \@sanitize@url \@href}%
\providecommand \@href[1]{\@@startlink{#1}\@@href}%
\providecommand \@@href[1]{\endgroup#1\@@endlink}%
\providecommand \@sanitize@url [0]{\catcode `\\12\catcode `\$12\catcode `\&12\catcode `\#12\catcode `\^12\catcode `\_12\catcode `\%12\relax}%
\providecommand \@@startlink[1]{}%
\providecommand \@@endlink[0]{}%
\providecommand \url  [0]{\begingroup\@sanitize@url \@url }%
\providecommand \@url [1]{\endgroup\@href {#1}{\urlprefix }}%
\providecommand \urlprefix  [0]{URL }%
\providecommand \Eprint [0]{\href }%
\providecommand \doibase [0]{https://doi.org/}%
\providecommand \selectlanguage [0]{\@gobble}%
\providecommand \bibinfo  [0]{\@secondoftwo}%
\providecommand \bibfield  [0]{\@secondoftwo}%
\providecommand \translation [1]{[#1]}%
\providecommand \BibitemOpen [0]{}%
\providecommand \bibitemStop [0]{}%
\providecommand \bibitemNoStop [0]{.\EOS\space}%
\providecommand \EOS [0]{\spacefactor3000\relax}%
\providecommand \BibitemShut  [1]{\csname bibitem#1\endcsname}%
\let\auto@bib@innerbib\@empty
\bibitem [{noa()}]{Fazio2024}%
  \BibitemOpen
  \href {https://www.scipost.org/SciPostPhysLectNotes.99} {\bibinfo {title} {{SciPost}: {SciPost} {Phys}. {Lect}. {Notes} 99 (2025) - {Many}-body open quantum systems}}\BibitemShut {NoStop}%
\bibitem [{\citenamefont {Kessler}\ \emph {et~al.}(2012)\citenamefont {Kessler}, \citenamefont {Giedke}, \citenamefont {Imamoglu}, \citenamefont {Yelin}, \citenamefont {Lukin},\ and\ \citenamefont {Cirac}}]{Kessler2012}%
  \BibitemOpen
  \bibfield  {author} {\bibinfo {author} {\bibfnamefont {E.~M.}\ \bibnamefont {Kessler}}, \bibinfo {author} {\bibfnamefont {G.}~\bibnamefont {Giedke}}, \bibinfo {author} {\bibfnamefont {A.}~\bibnamefont {Imamoglu}}, \bibinfo {author} {\bibfnamefont {S.~F.}\ \bibnamefont {Yelin}}, \bibinfo {author} {\bibfnamefont {M.~D.}\ \bibnamefont {Lukin}},\ and\ \bibinfo {author} {\bibfnamefont {J.~I.}\ \bibnamefont {Cirac}},\ }\bibfield  {title} {\bibinfo {title} {Dissipative phase transition in a central spin system},\ }\href {https://doi.org/10.1103/PhysRevA.86.012116} {\bibfield  {journal} {\bibinfo  {journal} {Phys. Rev. A}\ }\textbf {\bibinfo {volume} {86}},\ \bibinfo {pages} {012116} (\bibinfo {year} {2012})}\BibitemShut {NoStop}%
\bibitem [{\citenamefont {Minganti}\ \emph {et~al.}(2018)\citenamefont {Minganti}, \citenamefont {Biella}, \citenamefont {Bartolo},\ and\ \citenamefont {Ciuti}}]{Minganti2018}%
  \BibitemOpen
  \bibfield  {author} {\bibinfo {author} {\bibfnamefont {F.}~\bibnamefont {Minganti}}, \bibinfo {author} {\bibfnamefont {A.}~\bibnamefont {Biella}}, \bibinfo {author} {\bibfnamefont {N.}~\bibnamefont {Bartolo}},\ and\ \bibinfo {author} {\bibfnamefont {C.}~\bibnamefont {Ciuti}},\ }\bibfield  {title} {\bibinfo {title} {Spectral theory of liouvillians for dissipative phase transitions},\ }\href {https://doi.org/10.1103/PhysRevA.98.042118} {\bibfield  {journal} {\bibinfo  {journal} {Phys. Rev. A}\ }\textbf {\bibinfo {volume} {98}},\ \bibinfo {pages} {042118} (\bibinfo {year} {2018})}\BibitemShut {NoStop}%
\bibitem [{\citenamefont {Soriente}\ \emph {et~al.}(2021)\citenamefont {Soriente}, \citenamefont {Heugel}, \citenamefont {Omiya}, \citenamefont {Chitra},\ and\ \citenamefont {Zilberberg}}]{Soriente2021}%
  \BibitemOpen
  \bibfield  {author} {\bibinfo {author} {\bibfnamefont {M.}~\bibnamefont {Soriente}}, \bibinfo {author} {\bibfnamefont {T.~L.}\ \bibnamefont {Heugel}}, \bibinfo {author} {\bibfnamefont {K.}~\bibnamefont {Omiya}}, \bibinfo {author} {\bibfnamefont {R.}~\bibnamefont {Chitra}},\ and\ \bibinfo {author} {\bibfnamefont {O.}~\bibnamefont {Zilberberg}},\ }\bibfield  {title} {\bibinfo {title} {Distinctive class of dissipation-induced phase transitions and their universal characteristics},\ }\href {https://doi.org/10.1103/PhysRevResearch.3.023100} {\bibfield  {journal} {\bibinfo  {journal} {Phys. Rev. Res.}\ }\textbf {\bibinfo {volume} {3}},\ \bibinfo {pages} {023100} (\bibinfo {year} {2021})}\BibitemShut {NoStop}%
\bibitem [{\citenamefont {Debecker}\ \emph {et~al.}(2024)\citenamefont {Debecker}, \citenamefont {Martin},\ and\ \citenamefont {Damanet}}]{Damanet2024}%
  \BibitemOpen
  \bibfield  {author} {\bibinfo {author} {\bibfnamefont {B.}~\bibnamefont {Debecker}}, \bibinfo {author} {\bibfnamefont {J.}~\bibnamefont {Martin}},\ and\ \bibinfo {author} {\bibfnamefont {F.}~\bibnamefont {Damanet}},\ }\bibfield  {title} {\bibinfo {title} {Spectral theory of non-{M}arkovian dissipative phase transitions},\ }\href {https://doi.org/10.1103/PhysRevA.110.042201} {\bibfield  {journal} {\bibinfo  {journal} {Phys. Rev. A}\ }\textbf {\bibinfo {volume} {110}},\ \bibinfo {pages} {042201} (\bibinfo {year} {2024})}\BibitemShut {NoStop}%
\bibitem [{\citenamefont {Hannukainen}\ and\ \citenamefont {Larson}(2018)}]{Hannukainen2018}%
  \BibitemOpen
  \bibfield  {author} {\bibinfo {author} {\bibfnamefont {J.}~\bibnamefont {Hannukainen}}\ and\ \bibinfo {author} {\bibfnamefont {J.}~\bibnamefont {Larson}},\ }\bibfield  {title} {\bibinfo {title} {Dissipation-driven quantum phase transitions and symmetry breaking},\ }\href {https://doi.org/10.1103/PhysRevA.98.042113} {\bibfield  {journal} {\bibinfo  {journal} {Phys. Rev. A}\ }\textbf {\bibinfo {volume} {98}},\ \bibinfo {pages} {042113} (\bibinfo {year} {2018})}\BibitemShut {NoStop}%
\bibitem [{\citenamefont {Altland}\ \emph {et~al.}(2021)\citenamefont {Altland}, \citenamefont {Fleischhauer},\ and\ \citenamefont {Diehl}}]{Altland2021}%
  \BibitemOpen
  \bibfield  {author} {\bibinfo {author} {\bibfnamefont {A.}~\bibnamefont {Altland}}, \bibinfo {author} {\bibfnamefont {M.}~\bibnamefont {Fleischhauer}},\ and\ \bibinfo {author} {\bibfnamefont {S.}~\bibnamefont {Diehl}},\ }\bibfield  {title} {\bibinfo {title} {Symmetry classes of open fermionic quantum matter},\ }\href {https://doi.org/10.1103/PhysRevX.11.021037} {\bibfield  {journal} {\bibinfo  {journal} {Phys. Rev. X}\ }\textbf {\bibinfo {volume} {11}},\ \bibinfo {pages} {021037} (\bibinfo {year} {2021})}\BibitemShut {NoStop}%
\bibitem [{\citenamefont {Kawabata}\ \emph {et~al.}(2023)\citenamefont {Kawabata}, \citenamefont {Kulkarni}, \citenamefont {Li}, \citenamefont {Numasawa},\ and\ \citenamefont {Ryu}}]{Kawabata2023}%
  \BibitemOpen
  \bibfield  {author} {\bibinfo {author} {\bibfnamefont {K.}~\bibnamefont {Kawabata}}, \bibinfo {author} {\bibfnamefont {A.}~\bibnamefont {Kulkarni}}, \bibinfo {author} {\bibfnamefont {J.}~\bibnamefont {Li}}, \bibinfo {author} {\bibfnamefont {T.}~\bibnamefont {Numasawa}},\ and\ \bibinfo {author} {\bibfnamefont {S.}~\bibnamefont {Ryu}},\ }\bibfield  {title} {\bibinfo {title} {Symmetry of open quantum systems: Classification of dissipative quantum chaos},\ }\href {https://doi.org/10.1103/PRXQuantum.4.030328} {\bibfield  {journal} {\bibinfo  {journal} {PRX Quantum}\ }\textbf {\bibinfo {volume} {4}},\ \bibinfo {pages} {030328} (\bibinfo {year} {2023})}\BibitemShut {NoStop}%
\bibitem [{\citenamefont {S\'a}\ \emph {et~al.}(2023)\citenamefont {S\'a}, \citenamefont {Ribeiro},\ and\ \citenamefont {Prosen}}]{Sa2023}%
  \BibitemOpen
  \bibfield  {author} {\bibinfo {author} {\bibfnamefont {L.}~\bibnamefont {S\'a}}, \bibinfo {author} {\bibfnamefont {P.}~\bibnamefont {Ribeiro}},\ and\ \bibinfo {author} {\bibfnamefont {T.}~\bibnamefont {Prosen}},\ }\bibfield  {title} {\bibinfo {title} {Symmetry classification of many-body lindbladians: Tenfold way and beyond},\ }\href {https://doi.org/10.1103/PhysRevX.13.031019} {\bibfield  {journal} {\bibinfo  {journal} {Phys. Rev. X}\ }\textbf {\bibinfo {volume} {13}},\ \bibinfo {pages} {031019} (\bibinfo {year} {2023})}\BibitemShut {NoStop}%
\bibitem [{\citenamefont {Debecker}\ \emph {et~al.}(2025)\citenamefont {Debecker}, \citenamefont {Pausch}, \citenamefont {Louvet}, \citenamefont {Bastin}, \citenamefont {Martin},\ and\ \citenamefont {Damanet}}]{debecker_role_2025}%
  \BibitemOpen
  \bibfield  {author} {\bibinfo {author} {\bibfnamefont {B.}~\bibnamefont {Debecker}}, \bibinfo {author} {\bibfnamefont {L.}~\bibnamefont {Pausch}}, \bibinfo {author} {\bibfnamefont {J.}~\bibnamefont {Louvet}}, \bibinfo {author} {\bibfnamefont {T.}~\bibnamefont {Bastin}}, \bibinfo {author} {\bibfnamefont {J.}~\bibnamefont {Martin}},\ and\ \bibinfo {author} {\bibfnamefont {F.}~\bibnamefont {Damanet}},\ }\bibfield  {title} {\bibinfo {title} {Role of non-{Markovian} dissipation in quantum phase transitions: {Tricriticality}, spin squeezing, and directional symmetry breaking},\ }\href {https://doi.org/10.1103/4wq2-wyh4} {\bibfield  {journal} {\bibinfo  {journal} {Physical Review A}\ }\textbf {\bibinfo {volume} {112}},\ \bibinfo {pages} {012210} (\bibinfo {year} {2025})}\BibitemShut {NoStop}%
  \bibitem [{\citenamefont {Baumann}\ \emph {et~al.}(2010)\citenamefont {Baumann}, \citenamefont {Guerlin}, \citenamefont {Brennecke},\ and\ \citenamefont {Esslinger}}]{Baumann2010}%
  \BibitemOpen
  \bibfield  {author} {\bibinfo {author} {\bibfnamefont {K.}~\bibnamefont {Baumann}}, \bibinfo {author} {\bibfnamefont {C.}~\bibnamefont {Guerlin}}, \bibinfo {author} {\bibfnamefont {F.}~\bibnamefont {Brennecke}},\ and\ \bibinfo {author} {\bibfnamefont {T.}~\bibnamefont {Esslinger}},\ }\bibfield  {title} {\bibinfo {title} {Dicke quantum phase transition with a superfluid gas in an optical cavity},\ }\href {https://doi.org/10.1038/nature09009} {\bibfield  {journal} {\bibinfo  {journal} {Nature}\ }\textbf {\bibinfo {volume} {464}},\ \bibinfo {pages} {1301} (\bibinfo {year} {2010})}\BibitemShut {NoStop}%
\bibitem [{\citenamefont {Cai}\ \emph {et~al.}(2021)\citenamefont {Cai}, \citenamefont {Liu}, \citenamefont {Zhao}, \citenamefont {Wu}, \citenamefont {Mei}, \citenamefont {Jiang}, \citenamefont {He}, \citenamefont {Zhang}, \citenamefont {Zhou},\ and\ \citenamefont {Duan}}]{Cai2021}%
  \BibitemOpen
  \bibfield  {author} {\bibinfo {author} {\bibfnamefont {M.-L.}\ \bibnamefont {Cai}}, \bibinfo {author} {\bibfnamefont {Z.-D.}\ \bibnamefont {Liu}}, \bibinfo {author} {\bibfnamefont {W.-D.}\ \bibnamefont {Zhao}}, \bibinfo {author} {\bibfnamefont {Y.-K.}\ \bibnamefont {Wu}}, \bibinfo {author} {\bibfnamefont {Q.-X.}\ \bibnamefont {Mei}}, \bibinfo {author} {\bibfnamefont {Y.}~\bibnamefont {Jiang}}, \bibinfo {author} {\bibfnamefont {L.}~\bibnamefont {He}}, \bibinfo {author} {\bibfnamefont {X.}~\bibnamefont {Zhang}}, \bibinfo {author} {\bibfnamefont {Z.-C.}\ \bibnamefont {Zhou}},\ and\ \bibinfo {author} {\bibfnamefont {L.-M.}\ \bibnamefont {Duan}},\ }\bibfield  {title} {\bibinfo {title} {Observation of a quantum phase transition in the quantum rabi model with a single trapped ion},\ }\href@noop {} {\bibfield  {journal} {\bibinfo  {journal} {Nature communications}\ }\textbf {\bibinfo {volume} {12}},\ \bibinfo {pages} {1126} (\bibinfo {year} {2021})}\BibitemShut {NoStop}%
\bibitem [{\citenamefont {Chen}\ \emph {et~al.}(2023)\citenamefont {Chen}, \citenamefont {Fischer}, \citenamefont {Nojiri}, \citenamefont {Renger}, \citenamefont {Xie}, \citenamefont {Partanen}, \citenamefont {Pogorzalek}, \citenamefont {Fedorov}, \citenamefont {Marx}, \citenamefont {Deppe} \emph {et~al.}}]{Chen2023}%
  \BibitemOpen
  \bibfield  {author} {\bibinfo {author} {\bibfnamefont {Q.-M.}\ \bibnamefont {Chen}}, \bibinfo {author} {\bibfnamefont {M.}~\bibnamefont {Fischer}}, \bibinfo {author} {\bibfnamefont {Y.}~\bibnamefont {Nojiri}}, \bibinfo {author} {\bibfnamefont {M.}~\bibnamefont {Renger}}, \bibinfo {author} {\bibfnamefont {E.}~\bibnamefont {Xie}}, \bibinfo {author} {\bibfnamefont {M.}~\bibnamefont {Partanen}}, \bibinfo {author} {\bibfnamefont {S.}~\bibnamefont {Pogorzalek}}, \bibinfo {author} {\bibfnamefont {K.~G.}\ \bibnamefont {Fedorov}}, \bibinfo {author} {\bibfnamefont {A.}~\bibnamefont {Marx}}, \bibinfo {author} {\bibfnamefont {F.}~\bibnamefont {Deppe}}, \emph {et~al.},\ }\bibfield  {title} {\bibinfo {title} {Quantum behavior of the {D}uffing oscillator at the dissipative phase transition},\ }\href@noop {} {\bibfield  {journal} {\bibinfo  {journal} {Nature Communications}\ }\textbf {\bibinfo {volume} {14}},\ \bibinfo {pages} {2896} (\bibinfo {year} {2023})}\BibitemShut {NoStop}%
\bibitem [{\citenamefont {Rodriguez}\ \emph {et~al.}(2017)\citenamefont {Rodriguez}, \citenamefont {Casteels}, \citenamefont {Storme}, \citenamefont {Carlon~Zambon}, \citenamefont {Sagnes}, \citenamefont {Le~Gratiet}, \citenamefont {Galopin}, \citenamefont {Lema\^{\i}tre}, \citenamefont {Amo}, \citenamefont {Ciuti},\ and\ \citenamefont {Bloch}}]{Rodriguez2017}%
  \BibitemOpen
  \bibfield  {author} {\bibinfo {author} {\bibfnamefont {S.~R.~K.}\ \bibnamefont {Rodriguez}}, \bibinfo {author} {\bibfnamefont {W.}~\bibnamefont {Casteels}}, \bibinfo {author} {\bibfnamefont {F.}~\bibnamefont {Storme}}, \bibinfo {author} {\bibfnamefont {N.}~\bibnamefont {Carlon~Zambon}}, \bibinfo {author} {\bibfnamefont {I.}~\bibnamefont {Sagnes}}, \bibinfo {author} {\bibfnamefont {L.}~\bibnamefont {Le~Gratiet}}, \bibinfo {author} {\bibfnamefont {E.}~\bibnamefont {Galopin}}, \bibinfo {author} {\bibfnamefont {A.}~\bibnamefont {Lema\^{\i}tre}}, \bibinfo {author} {\bibfnamefont {A.}~\bibnamefont {Amo}}, \bibinfo {author} {\bibfnamefont {C.}~\bibnamefont {Ciuti}},\ and\ \bibinfo {author} {\bibfnamefont {J.}~\bibnamefont {Bloch}},\ }\bibfield  {title} {\bibinfo {title} {Probing a dissipative phase transition via dynamical optical hysteresis},\ }\href {https://doi.org/10.1103/PhysRevLett.118.247402} {\bibfield  {journal} {\bibinfo  {journal} {Phys. Rev. Lett.}\ }\textbf {\bibinfo {volume} {118}},\ \bibinfo {pages}
  {247402} (\bibinfo {year} {2017})}\BibitemShut {NoStop}%
\bibitem [{\citenamefont {Beaulieu}\ \emph {et~al.}(2025)\citenamefont {Beaulieu}, \citenamefont {Minganti}, \citenamefont {Frasca}, \citenamefont {Savona}, \citenamefont {Felicetti}, \citenamefont {Di~Candia},\ and\ \citenamefont {Scarlino}}]{Beaulieu2025}%
  \BibitemOpen
  \bibfield  {author} {\bibinfo {author} {\bibfnamefont {G.}~\bibnamefont {Beaulieu}}, \bibinfo {author} {\bibfnamefont {F.}~\bibnamefont {Minganti}}, \bibinfo {author} {\bibfnamefont {S.}~\bibnamefont {Frasca}}, \bibinfo {author} {\bibfnamefont {V.}~\bibnamefont {Savona}}, \bibinfo {author} {\bibfnamefont {S.}~\bibnamefont {Felicetti}}, \bibinfo {author} {\bibfnamefont {R.}~\bibnamefont {Di~Candia}},\ and\ \bibinfo {author} {\bibfnamefont {P.}~\bibnamefont {Scarlino}},\ }\bibfield  {title} {\bibinfo {title} {Observation of first-and second-order dissipative phase transitions in a two-photon driven {K}err resonator},\ }\href@noop {} {\bibfield  {journal} {\bibinfo  {journal} {Nature Communications}\ }\textbf {\bibinfo {volume} {16}},\ \bibinfo {pages} {1954} (\bibinfo {year} {2025})}\BibitemShut {NoStop}%
\bibitem [{\citenamefont {Verstraete}\ \emph {et~al.}(2009)\citenamefont {Verstraete}, \citenamefont {Wolf},\ and\ \citenamefont {Ignacio~Cirac}}]{verstraete_quantum_2009}%
  \BibitemOpen
  \bibfield  {author} {\bibinfo {author} {\bibfnamefont {F.}~\bibnamefont {Verstraete}}, \bibinfo {author} {\bibfnamefont {M.~M.}\ \bibnamefont {Wolf}},\ and\ \bibinfo {author} {\bibfnamefont {J.}~\bibnamefont {Ignacio~Cirac}},\ }\bibfield  {title} {{\bibinfo {title} {Quantum computation and quantum-state engineering driven by dissipation}},\ }\href {https://doi.org/10.1038/nphys1342} {\bibfield  {journal} {\bibinfo  {journal} {Nature Physics}\ }\textbf {\bibinfo {volume} {5}},\ \bibinfo {pages} {633} (\bibinfo {year} {2009})}\BibitemShut {NoStop}%
  \bibitem [{\citenamefont {Mirrahimi}\ \emph {et~al.}(2014)\citenamefont {Mirrahimi}, \citenamefont {Leghtas}, \citenamefont {Albert}, \citenamefont {Touzard}, \citenamefont {Schoelkopf}, \citenamefont {Jiang},\ and\ \citenamefont {Devoret}}]{mirrahimi_dynamically_2014}%
  \BibitemOpen
  \bibfield  {author} {\bibinfo {author} {\bibfnamefont {M.}~\bibnamefont {Mirrahimi}}, \bibinfo {author} {\bibfnamefont {Z.}~\bibnamefont {Leghtas}}, \bibinfo {author} {\bibfnamefont {V.~V.}\ \bibnamefont {Albert}}, \bibinfo {author} {\bibfnamefont {S.}~\bibnamefont {Touzard}}, \bibinfo {author} {\bibfnamefont {R.~J.}\ \bibnamefont {Schoelkopf}}, \bibinfo {author} {\bibfnamefont {L.}~\bibnamefont {Jiang}},\ and\ \bibinfo {author} {\bibfnamefont {M.~H.}\ \bibnamefont {Devoret}},\ }\bibfield  {title} {{\bibinfo {title} {Dynamically protected cat-qubits: a new paradigm for universal quantum computation}},\ }\href {https://doi.org/10.1088/1367-2630/16/4/045014} {\bibfield  {journal} {\bibinfo  {journal} {New Journal of Physics}\ }\textbf {\bibinfo {volume} {16}},\ \bibinfo {pages} {045014} (\bibinfo {year} {2014})}\BibitemShut {NoStop}%
\bibitem [{\citenamefont {Ding}\ \emph {et~al.}(2022)\citenamefont {Ding}, \citenamefont {Liu}, \citenamefont {Shi}, \citenamefont {Guo}, \citenamefont {Mølmer},\ and\ \citenamefont {Adams}}]{ding_enhanced_2022}%
  \BibitemOpen
  \bibfield  {author} {\bibinfo {author} {\bibfnamefont {D.-S.}\ \bibnamefont {Ding}}, \bibinfo {author} {\bibfnamefont {Z.-K.}\ \bibnamefont {Liu}}, \bibinfo {author} {\bibfnamefont {B.-S.}\ \bibnamefont {Shi}}, \bibinfo {author} {\bibfnamefont {G.-C.}\ \bibnamefont {Guo}}, \bibinfo {author} {\bibfnamefont {K.}~\bibnamefont {Mølmer}},\ and\ \bibinfo {author} {\bibfnamefont {C.~S.}\ \bibnamefont {Adams}},\ }\bibfield  {title} {{\bibinfo {title} {Enhanced metrology at the critical point of a many-body {Rydberg} atomic system}},\ }\href {https://doi.org/10.1038/s41567-022-01777-8} {\bibfield  {journal} {\bibinfo  {journal} {Nature Physics}\ }\textbf {\bibinfo {volume} {18}},\ \bibinfo {pages} {1447} (\bibinfo {year} {2022})}\BibitemShut {NoStop}%
\bibitem [{\citenamefont {Salvatori}\ \emph {et~al.}(2014)\citenamefont {Salvatori}, \citenamefont {Mandarino},\ and\ \citenamefont {Paris}}]{salvatori_quantum_2014}%
  \BibitemOpen
  \bibfield  {author} {\bibinfo {author} {\bibfnamefont {G.}~\bibnamefont {Salvatori}}, \bibinfo {author} {\bibfnamefont {A.}~\bibnamefont {Mandarino}},\ and\ \bibinfo {author} {\bibfnamefont {M.~G.~A.}\ \bibnamefont {Paris}},\ }\bibfield  {title} {\bibinfo {title} {Quantum metrology in {Lipkin}-{Meshkov}-{Glick} critical systems},\ }\href {https://doi.org/10.1103/PhysRevA.90.022111} {\bibfield  {journal} {\bibinfo  {journal} {Physical Review A}\ }\textbf {\bibinfo {volume} {90}},\ \bibinfo {pages} {022111} (\bibinfo {year} {2014})}\BibitemShut {NoStop}%
\bibitem [{\citenamefont {Ma}\ \emph {et~al.}(2017)\citenamefont {Ma}, \citenamefont {Su},\ and\ \citenamefont {Sun}}]{ma_quantum_2017}%
  \BibitemOpen
  \bibfield  {author} {\bibinfo {author} {\bibfnamefont {Y.-H.}\ \bibnamefont {Ma}}, \bibinfo {author} {\bibfnamefont {S.-H.}\ \bibnamefont {Su}},\ and\ \bibinfo {author} {\bibfnamefont {C.-P.}\ \bibnamefont {Sun}},\ }\bibfield  {title} {\bibinfo {title} {Quantum thermodynamic cycle with quantum phase transition},\ }\href {https://doi.org/10.1103/PhysRevE.96.022143} {\bibfield  {journal} {\bibinfo  {journal} {Physical Review E}\ }\textbf {\bibinfo {volume} {96}},\ \bibinfo {pages} {022143} (\bibinfo {year} {2017})}\BibitemShut {NoStop}%
\bibitem [{\citenamefont {Garraway}(2011)}]{Garraway2011}%
  \BibitemOpen
  \bibfield  {author} {\bibinfo {author} {\bibfnamefont {B.~M.}\ \bibnamefont {Garraway}},\ }\bibfield  {title} {\bibinfo {title} {The dicke model in quantum optics: Dicke model revisited},\ }\href@noop {} {\bibfield  {journal} {\bibinfo  {journal} {Philosophical Transactions of the Royal Society A: Mathematical, Physical and Engineering Sciences}\ }\textbf {\bibinfo {volume} {369}},\ \bibinfo {pages} {1137} (\bibinfo {year} {2011})}\BibitemShut {NoStop}%
\bibitem [{\citenamefont {Kirton}\ \emph {et~al.}(2019)\citenamefont {Kirton}, \citenamefont {Roses}, \citenamefont {Keeling},\ and\ \citenamefont {Dalla~Torre}}]{Kirton2019}%
  \BibitemOpen
  \bibfield  {author} {\bibinfo {author} {\bibfnamefont {P.}~\bibnamefont {Kirton}}, \bibinfo {author} {\bibfnamefont {M.~M.}\ \bibnamefont {Roses}}, \bibinfo {author} {\bibfnamefont {J.}~\bibnamefont {Keeling}},\ and\ \bibinfo {author} {\bibfnamefont {E.~G.}\ \bibnamefont {Dalla~Torre}},\ }\bibfield  {title} {\bibinfo {title} {Introduction to the {D}icke model: From equilibrium to nonequilibrium, and vice versa},\ }\href {https://doi.org/https://doi.org/10.1002/qute.201800043} {\bibfield  {journal} {\bibinfo  {journal} {Advanced Quantum Technologies}\ }\textbf {\bibinfo {volume} {2}},\ \bibinfo {pages} {1800043} (\bibinfo {year} {2019})}\BibitemShut {NoStop}%
\bibitem [{\citenamefont {Morrison}\ and\ \citenamefont {Parkins}(2008{\natexlab{a}})}]{Morrison2008}%
  \BibitemOpen
  \bibfield  {author} {\bibinfo {author} {\bibfnamefont {S.}~\bibnamefont {Morrison}}\ and\ \bibinfo {author} {\bibfnamefont {A.~S.}\ \bibnamefont {Parkins}},\ }\bibfield  {title} {\bibinfo {title} {Dynamical quantum phase transitions in the dissipative {L}ipkin-{M}eshkov-{G}lick model with proposed realization in optical cavity {QED}},\ }\href {https://doi.org/10.1103/PhysRevLett.100.040403} {\bibfield  {journal} {\bibinfo  {journal} {Phys. Rev. Lett.}\ }\textbf {\bibinfo {volume} {100}},\ \bibinfo {pages} {040403} (\bibinfo {year} {2008}{\natexlab{a}})}\BibitemShut {NoStop}%
\bibitem [{\citenamefont {Morrison}\ and\ \citenamefont {Parkins}(2008{\natexlab{b}})}]{Morrison2008a}%
  \BibitemOpen
  \bibfield  {author} {\bibinfo {author} {\bibfnamefont {S.}~\bibnamefont {Morrison}}\ and\ \bibinfo {author} {\bibfnamefont {A.~S.}\ \bibnamefont {Parkins}},\ }\bibfield  {title} {\bibinfo {title} {Collective spin systems in dispersive optical cavity {QED}: Quantum phase transitions and entanglement},\ }\href {https://doi.org/10.1103/PhysRevA.77.043810} {\bibfield  {journal} {\bibinfo  {journal} {Phys. Rev. A}\ }\textbf {\bibinfo {volume} {77}},\ \bibinfo {pages} {043810} (\bibinfo {year} {2008}{\natexlab{b}})}\BibitemShut {NoStop}%
\bibitem [{\citenamefont {Ferreira}\ and\ \citenamefont {Ribeiro}(2019)}]{Ferreira2019}%
  \BibitemOpen
  \bibfield  {author} {\bibinfo {author} {\bibfnamefont {J.~S.}\ \bibnamefont {Ferreira}}\ and\ \bibinfo {author} {\bibfnamefont {P.}~\bibnamefont {Ribeiro}},\ }\bibfield  {title} {\bibinfo {title} {Lipkin-{M}eshkov-{G}lick model with markovian dissipation: A description of a collective spin on a metallic surface},\ }\href {https://doi.org/10.1103/PhysRevB.100.184422} {\bibfield  {journal} {\bibinfo  {journal} {Phys. Rev. B}\ }\textbf {\bibinfo {volume} {100}},\ \bibinfo {pages} {184422} (\bibinfo {year} {2019})}\BibitemShut {NoStop}%
\bibitem [{\citenamefont {Scully}\ and\ \citenamefont {Zubairy}(1997)}]{Scully1997}%
  \BibitemOpen
  \bibfield  {author} {\bibinfo {author} {\bibfnamefont {M.~O.}\ \bibnamefont {Scully}}\ and\ \bibinfo {author} {\bibfnamefont {M.~S.}\ \bibnamefont {Zubairy}},\ }\href@noop {} {\emph {\bibinfo {title} {Quantum Optics}}}\ (\bibinfo  {publisher} {Cambridge University Press},\ \bibinfo {year} {1997})\BibitemShut {NoStop}%
\bibitem [{\citenamefont {Carollo}\ \emph {et~al.}(2019)\citenamefont {Carollo}, \citenamefont {Gillman}, \citenamefont {Weimer},\ and\ \citenamefont {Lesanovsky}}]{Carollo2019}%
  \BibitemOpen
  \bibfield  {author} {\bibinfo {author} {\bibfnamefont {F.}~\bibnamefont {Carollo}}, \bibinfo {author} {\bibfnamefont {E.}~\bibnamefont {Gillman}}, \bibinfo {author} {\bibfnamefont {H.}~\bibnamefont {Weimer}},\ and\ \bibinfo {author} {\bibfnamefont {I.}~\bibnamefont {Lesanovsky}},\ }\bibfield  {title} {\bibinfo {title} {Critical behavior of the quantum contact process in one dimension},\ }\href {https://doi.org/10.1103/PhysRevLett.123.100604} {\bibfield  {journal} {\bibinfo  {journal} {Phys. Rev. Lett.}\ }\textbf {\bibinfo {volume} {123}},\ \bibinfo {pages} {100604} (\bibinfo {year} {2019})}\BibitemShut {NoStop}%
\bibitem [{\citenamefont {Gillman}\ \emph {et~al.}(2019)\citenamefont {Gillman}, \citenamefont {Carollo},\ and\ \citenamefont {Lesanovsky}}]{Gillman2019}%
  \BibitemOpen
  \bibfield  {author} {\bibinfo {author} {\bibfnamefont {E.}~\bibnamefont {Gillman}}, \bibinfo {author} {\bibfnamefont {F.}~\bibnamefont {Carollo}},\ and\ \bibinfo {author} {\bibfnamefont {I.}~\bibnamefont {Lesanovsky}},\ }\bibfield  {title} {\bibinfo {title} {Numerical simulation of critical dissipative non-equilibrium quantum systems with an absorbing state},\ }\href {https://doi.org/10.1088/1367-2630/ab43b0} {\bibfield  {journal} {\bibinfo  {journal} {New Journal of Physics}\ }\textbf {\bibinfo {volume} {21}},\ \bibinfo {pages} {093064} (\bibinfo {year} {2019})}\BibitemShut {NoStop}%
\bibitem [{\citenamefont {Carollo}\ and\ \citenamefont {Lesanovsky}(2022)}]{Carollo2022}%
  \BibitemOpen
  \bibfield  {author} {\bibinfo {author} {\bibfnamefont {F.}~\bibnamefont {Carollo}}\ and\ \bibinfo {author} {\bibfnamefont {I.}~\bibnamefont {Lesanovsky}},\ }\bibfield  {title} {\bibinfo {title} {Nonequilibrium dark space phase transition},\ }\href {https://doi.org/10.1103/PhysRevLett.128.040603} {\bibfield  {journal} {\bibinfo  {journal} {Phys. Rev. Lett.}\ }\textbf {\bibinfo {volume} {128}},\ \bibinfo {pages} {040603} (\bibinfo {year} {2022})}\BibitemShut {NoStop}%
\bibitem [{\citenamefont {Iemini}\ \emph {et~al.}(2018)\citenamefont {Iemini}, \citenamefont {Russomanno}, \citenamefont {Keeling}, \citenamefont {Schir\`o}, \citenamefont {Dalmonte},\ and\ \citenamefont {Fazio}}]{Iemini2018}%
  \BibitemOpen
  \bibfield  {author} {\bibinfo {author} {\bibfnamefont {F.}~\bibnamefont {Iemini}}, \bibinfo {author} {\bibfnamefont {A.}~\bibnamefont {Russomanno}}, \bibinfo {author} {\bibfnamefont {J.}~\bibnamefont {Keeling}}, \bibinfo {author} {\bibfnamefont {M.}~\bibnamefont {Schir\`o}}, \bibinfo {author} {\bibfnamefont {M.}~\bibnamefont {Dalmonte}},\ and\ \bibinfo {author} {\bibfnamefont {R.}~\bibnamefont {Fazio}},\ }\bibfield  {title} {\bibinfo {title} {Boundary time crystals},\ }\href {https://doi.org/10.1103/PhysRevLett.121.035301} {\bibfield  {journal} {\bibinfo  {journal} {Phys. Rev. Lett.}\ }\textbf {\bibinfo {volume} {121}},\ \bibinfo {pages} {035301} (\bibinfo {year} {2018})}\BibitemShut {NoStop}%
\bibitem [{\citenamefont {Prosen}(2012{\natexlab{a}})}]{Prosen2012}%
  \BibitemOpen
  \bibfield  {author} {\bibinfo {author} {\bibfnamefont {T.}~\bibnamefont {Prosen}},\ }\bibfield  {title} {\bibinfo {title} {$\mathbb{P}\mathbb{T}$-symmetric quantum liouvillean dynamics},\ }\href {https://doi.org/10.1103/PhysRevLett.109.090404} {\bibfield  {journal} {\bibinfo  {journal} {Phys. Rev. Lett.}\ }\textbf {\bibinfo {volume} {109}},\ \bibinfo {pages} {090404} (\bibinfo {year} {2012}{\natexlab{a}})}\BibitemShut {NoStop}%
\bibitem [{\citenamefont {Prosen}(2012{\natexlab{b}})}]{Prosen2012a}%
  \BibitemOpen
  \bibfield  {author} {\bibinfo {author} {\bibfnamefont {T.}~\bibnamefont {Prosen}},\ }\bibfield  {title} {\bibinfo {title} {Generic examples of $\mathbb{P}\mathbb{T}$-symmetric qubit (spin-1/2) liouvillian dynamics},\ }\href {https://doi.org/10.1103/PhysRevA.86.044103} {\bibfield  {journal} {\bibinfo  {journal} {Phys. Rev. A}\ }\textbf {\bibinfo {volume} {86}},\ \bibinfo {pages} {044103} (\bibinfo {year} {2012}{\natexlab{b}})}\BibitemShut {NoStop}%
\bibitem [{\citenamefont {Huber}\ \emph {et~al.}(2020{\natexlab{a}})\citenamefont {Huber}, \citenamefont {Kirton}, \citenamefont {Rotter},\ and\ \citenamefont {Rabl}}]{Huber2020}%
  \BibitemOpen
  \bibfield  {author} {\bibinfo {author} {\bibfnamefont {J.}~\bibnamefont {Huber}}, \bibinfo {author} {\bibfnamefont {P.}~\bibnamefont {Kirton}}, \bibinfo {author} {\bibfnamefont {S.}~\bibnamefont {Rotter}},\ and\ \bibinfo {author} {\bibfnamefont {P.}~\bibnamefont {Rabl}},\ }\bibfield  {title} {\bibinfo {title} {{Emergence of PT-symmetry breaking in open quantum systems}},\ }\href {https://doi.org/10.21468/SciPostPhys.9.4.052} {\bibfield  {journal} {\bibinfo  {journal} {SciPost Phys.}\ }\textbf {\bibinfo {volume} {9}},\ \bibinfo {pages} {052} (\bibinfo {year} {2020}{\natexlab{a}})}\BibitemShut {NoStop}%
\bibitem [{\citenamefont {Huber}\ \emph {et~al.}(2020{\natexlab{b}})\citenamefont {Huber}, \citenamefont {Kirton},\ and\ \citenamefont {Rabl}}]{Huber2020a}%
  \BibitemOpen
  \bibfield  {author} {\bibinfo {author} {\bibfnamefont {J.}~\bibnamefont {Huber}}, \bibinfo {author} {\bibfnamefont {P.}~\bibnamefont {Kirton}},\ and\ \bibinfo {author} {\bibfnamefont {P.}~\bibnamefont {Rabl}},\ }\bibfield  {title} {\bibinfo {title} {Nonequilibrium magnetic phases in spin lattices with gain and loss},\ }\href {https://doi.org/10.1103/PhysRevA.102.012219} {\bibfield  {journal} {\bibinfo  {journal} {Phys. Rev. A}\ }\textbf {\bibinfo {volume} {102}},\ \bibinfo {pages} {012219} (\bibinfo {year} {2020}{\natexlab{b}})}\BibitemShut {NoStop}%
\bibitem [{\citenamefont {Nakanishi}\ and\ \citenamefont {Sasamoto}(2022)}]{Nakanishi2022}%
  \BibitemOpen
  \bibfield  {author} {\bibinfo {author} {\bibfnamefont {Y.}~\bibnamefont {Nakanishi}}\ and\ \bibinfo {author} {\bibfnamefont {T.}~\bibnamefont {Sasamoto}},\ }\bibfield  {title} {\bibinfo {title} {$\mathcal{PT}$ phase transition in open quantum systems with lindblad dynamics},\ }\href {https://doi.org/10.1103/PhysRevA.105.022219} {\bibfield  {journal} {\bibinfo  {journal} {Phys. Rev. A}\ }\textbf {\bibinfo {volume} {105}},\ \bibinfo {pages} {022219} (\bibinfo {year} {2022})}\BibitemShut {NoStop}%
\bibitem [{\citenamefont {Ashida}\ \emph {et~al.}(2020)\citenamefont {Ashida}, \citenamefont {Gong},\ and\ \citenamefont {Ueda}}]{Ashida2020}%
  \BibitemOpen
  \bibfield  {author} {\bibinfo {author} {\bibfnamefont {Y.}~\bibnamefont {Ashida}}, \bibinfo {author} {\bibfnamefont {Z.}~\bibnamefont {Gong}},\ and\ \bibinfo {author} {\bibfnamefont {M.}~\bibnamefont {Ueda}},\ }\bibfield  {title} {\bibinfo {title} {Non-hermitian physics},\ }\href@noop {} {\bibfield  {journal} {\bibinfo  {journal} {Advances in Physics}\ }\textbf {\bibinfo {volume} {69}},\ \bibinfo {pages} {249} (\bibinfo {year} {2020})}\BibitemShut {NoStop}%
\bibitem [{\citenamefont {Arkhipov}\ \emph {et~al.}(2020)\citenamefont {Arkhipov}, \citenamefont {Miranowicz}, \citenamefont {Minganti},\ and\ \citenamefont {Nori}}]{Arkhipov2020}%
  \BibitemOpen
  \bibfield  {author} {\bibinfo {author} {\bibfnamefont {I.~I.}\ \bibnamefont {Arkhipov}}, \bibinfo {author} {\bibfnamefont {A.}~\bibnamefont {Miranowicz}}, \bibinfo {author} {\bibfnamefont {F.}~\bibnamefont {Minganti}},\ and\ \bibinfo {author} {\bibfnamefont {F.}~\bibnamefont {Nori}},\ }\bibfield  {title} {\bibinfo {title} {Quantum and semiclassical exceptional points of a linear system of coupled cavities with losses and gain within the scully-lamb laser theory},\ }\href {https://doi.org/10.1103/PhysRevA.101.013812} {\bibfield  {journal} {\bibinfo  {journal} {Phys. Rev. A}\ }\textbf {\bibinfo {volume} {101}},\ \bibinfo {pages} {013812} (\bibinfo {year} {2020})}\BibitemShut {NoStop}%
\bibitem [{\citenamefont {Lipkin}\ \emph {et~al.}(1965)\citenamefont {Lipkin}, \citenamefont {Meshkov},\ and\ \citenamefont {Glick}}]{Lipkin1965}%
  \BibitemOpen
  \bibfield  {author} {\bibinfo {author} {\bibfnamefont {H.}~\bibnamefont {Lipkin}}, \bibinfo {author} {\bibfnamefont {N.}~\bibnamefont {Meshkov}},\ and\ \bibinfo {author} {\bibfnamefont {A.}~\bibnamefont {Glick}},\ }\bibfield  {title} {\bibinfo {title} {Validity of many-body approximation methods for a solvable model: (i). exact solutions and perturbation theory},\ }\href {https://doi.org/https://doi.org/10.1016/0029-5582(65)90862-X} {\bibfield  {journal} {\bibinfo  {journal} {Nuclear Physics}\ }\textbf {\bibinfo {volume} {62}},\ \bibinfo {pages} {188} (\bibinfo {year} {1965})}\BibitemShut {NoStop}%
\bibitem [{\citenamefont {Fiorelli}\ \emph {et~al.}(2023)\citenamefont {Fiorelli}, \citenamefont {Müller}, \citenamefont {Lesanovsky},\ and\ \citenamefont {Carollo}}]{Fiorelli2023}%
  \BibitemOpen
  \bibfield  {author} {\bibinfo {author} {\bibfnamefont {E.}~\bibnamefont {Fiorelli}}, \bibinfo {author} {\bibfnamefont {M.}~\bibnamefont {Müller}}, \bibinfo {author} {\bibfnamefont {I.}~\bibnamefont {Lesanovsky}},\ and\ \bibinfo {author} {\bibfnamefont {F.}~\bibnamefont {Carollo}},\ }\bibfield  {title} {\bibinfo {title} {Mean-field dynamics of open quantum systems with collective operator-valued rates: validity and application},\ }\href {https://doi.org/10.1088/1367-2630/ace470} {\bibfield  {journal} {\bibinfo  {journal} {New Journal of Physics}\ }\textbf {\bibinfo {volume} {25}},\ \bibinfo {pages} {083010} (\bibinfo {year} {2023})}\BibitemShut {NoStop}%
\bibitem [{\citenamefont {Fowler-Wright}\ \emph {et~al.}(2023)\citenamefont {Fowler-Wright}, \citenamefont {Arnard\'ottir}, \citenamefont {Kirton}, \citenamefont {Lovett},\ and\ \citenamefont {Keeling}}]{Wright2023}%
  \BibitemOpen
  \bibfield  {author} {\bibinfo {author} {\bibfnamefont {P.}~\bibnamefont {Fowler-Wright}}, \bibinfo {author} {\bibfnamefont {K.~B.}\ \bibnamefont {Arnard\'ottir}}, \bibinfo {author} {\bibfnamefont {P.}~\bibnamefont {Kirton}}, \bibinfo {author} {\bibfnamefont {B.~W.}\ \bibnamefont {Lovett}},\ and\ \bibinfo {author} {\bibfnamefont {J.}~\bibnamefont {Keeling}},\ }\bibfield  {title} {\bibinfo {title} {Determining the validity of cumulant expansions for central spin models},\ }\href {https://doi.org/10.1103/PhysRevResearch.5.033148} {\bibfield  {journal} {\bibinfo  {journal} {Phys. Rev. Res.}\ }\textbf {\bibinfo {volume} {5}},\ \bibinfo {pages} {033148} (\bibinfo {year} {2023})}\BibitemShut {NoStop}%
\bibitem [{Note1()}]{Note1}%
  \BibitemOpen
  \bibinfo {note} {Note that this is different to the condition imposed by the conservation of total angular momentum $\langle (S_A^x)^2\rangle + \langle (S_A^y)^2\rangle + \langle (S_A^z)^2\rangle = S(S+1)$ which is true for any state.}\BibitemShut {Stop}%
\bibitem [{\citenamefont {Tilma}\ \emph {et~al.}(2016)\citenamefont {Tilma}, \citenamefont {Everitt}, \citenamefont {Samson}, \citenamefont {Munro},\ and\ \citenamefont {Nemoto}}]{Tilma2016}%
  \BibitemOpen
  \bibfield  {author} {\bibinfo {author} {\bibfnamefont {T.}~\bibnamefont {Tilma}}, \bibinfo {author} {\bibfnamefont {M.~J.}\ \bibnamefont {Everitt}}, \bibinfo {author} {\bibfnamefont {J.~H.}\ \bibnamefont {Samson}}, \bibinfo {author} {\bibfnamefont {W.~J.}\ \bibnamefont {Munro}},\ and\ \bibinfo {author} {\bibfnamefont {K.}~\bibnamefont {Nemoto}},\ }\bibfield  {title} {\bibinfo {title} {Wigner functions for arbitrary quantum systems},\ }\href {https://doi.org/10.1103/PhysRevLett.117.180401} {\bibfield  {journal} {\bibinfo  {journal} {Phys. Rev. Lett.}\ }\textbf {\bibinfo {volume} {117}},\ \bibinfo {pages} {180401} (\bibinfo {year} {2016})}\BibitemShut {NoStop}%
  \bibitem [{\citenamefont {Benettin}\ \emph {et~al.}(1980)\citenamefont {Benettin}, \citenamefont {Galgani}, \citenamefont {Giorgilli},\ and\ \citenamefont {Strelcyn}}]{benettin_lyapunov_1980}%
  \BibitemOpen
  \bibfield  {author} {\bibinfo {author} {\bibfnamefont {G.}~\bibnamefont {Benettin}}, \bibinfo {author} {\bibfnamefont {L.}~\bibnamefont {Galgani}}, \bibinfo {author} {\bibfnamefont {A.}~\bibnamefont {Giorgilli}},\ and\ \bibinfo {author} {\bibfnamefont {J.-M.}\ \bibnamefont {Strelcyn}},\ }\bibfield  {title} {{\bibinfo {title} {Lyapunov {Characteristic} {Exponents} for smooth dynamical systems and for {H}amiltonian systems; a method for computing all of them. {Part} 1: {Theory}}},\ }\href{https://doi.org/10.1007/BF02128236} {\bibfield  {journal} {\bibinfo  {journal} {Meccanica}\ }\textbf {\bibinfo {volume} {15}},\ \bibinfo {pages} {9} (\bibinfo {year} {1980})}\BibitemShut {NoStop}%
\bibitem [{\citenamefont {Geist}\ \emph {et~al.}(1990)\citenamefont {Geist}, \citenamefont {Parlitz},\ and\ \citenamefont {Lauterborn}}]{geist_comparison_1990}%
  \BibitemOpen
  \bibfield  {author} {\bibinfo {author} {\bibfnamefont {K.}~\bibnamefont {Geist}}, \bibinfo {author} {\bibfnamefont {U.}~\bibnamefont {Parlitz}},\ and\ \bibinfo {author} {\bibfnamefont {W.}~\bibnamefont {Lauterborn}},\ }\bibfield  {title} {\bibinfo {title} {Comparison of {Different} {Methods} for {Computing} {Lyapunov} {Exponents}},\ }\href {https://doi.org/10.1143/PTP.83.875} {\bibfield  {journal} {\bibinfo  {journal} {Progress of Theoretical Physics}\ }\textbf {\bibinfo {volume} {83}},\ \bibinfo {pages} {875} (\bibinfo {year} {1990})}\BibitemShut {NoStop}%
  \bibitem [{\citenamefont {Datseris}(2018)}]{datseris_dynamicalsystemsjl_2018}%
  \BibitemOpen
  \bibfield  {author} {\bibinfo {author} {\bibfnamefont {G.}~\bibnamefont {Datseris}},\ }\bibfield  {title} {{\bibinfo {title} {{DynamicalSystems}.jl: {A} {Julia} software library for chaos and nonlinear dynamics}},\ }\href {https://doi.org/10.21105/joss.00598} {\bibfield  {journal} {\bibinfo  {journal} {Journal of Open Source Software}\ }\textbf {\bibinfo {volume} {3}},\ \bibinfo {pages} {598} (\bibinfo {year} {2018})}\BibitemShut {NoStop}%
  \bibitem [{\citenamefont {Solanki}\ and\ \citenamefont {Minganti}(2025)}]{solanki_chaos_2025}%
  \BibitemOpen
  \bibfield  {author} {\bibinfo {author} {\bibfnamefont {P.}~\bibnamefont {Solanki}}\ and\ \bibinfo {author} {\bibfnamefont {F.}~\bibnamefont {Minganti}},\ }\bibfield  {title} {\bibinfo {title} {Chaos as a manifestation of time-translation symmetry breaking},\ }\href {https://doi.org/10.1103/b6gq-z5nf} {\bibfield  {journal} {\bibinfo  {journal} {Physical Review B}\ }\textbf {\bibinfo {volume} {112}},\ \bibinfo {pages} {134311} (\bibinfo {year} {2025})}\BibitemShut {NoStop}%
\bibitem [{\citenamefont {Mu\~noz Arias}\ \emph {et~al.}(2023)\citenamefont {Mu\~noz Arias}, \citenamefont {Deutsch},\ and\ \citenamefont {Poggi}}]{Munoz2023}%
  \BibitemOpen
  \bibfield  {author} {\bibinfo {author} {\bibfnamefont {M.~H.}\ \bibnamefont {Mu\~noz Arias}}, \bibinfo {author} {\bibfnamefont {I.~H.}\ \bibnamefont {Deutsch}},\ and\ \bibinfo {author} {\bibfnamefont {P.~M.}\ \bibnamefont {Poggi}},\ }\bibfield  {title} {\bibinfo {title} {Phase-space geometry and optimal state preparation in quantum metrology with collective spins},\ }\href {https://doi.org/10.1103/PRXQuantum.4.020314} {\bibfield  {journal} {\bibinfo  {journal} {PRX Quantum}\ }\textbf {\bibinfo {volume} {4}},\ \bibinfo {pages} {020314} (\bibinfo {year} {2023})}\BibitemShut {NoStop}%
\bibitem [{\citenamefont {Champion}\ \emph {et~al.}(2025)\citenamefont {Champion}, \citenamefont {Schwartz}, \citenamefont {Ijaz}, \citenamefont {Xu}, \citenamefont {Campbell}, \citenamefont {Landi},\ and\ \citenamefont {Blok}}]{champion_analog_2025}%
  \BibitemOpen
  \bibfield  {author} {\bibinfo {author} {\bibfnamefont {E.}~\bibnamefont {Champion}}, \bibinfo {author} {\bibfnamefont {A.}~\bibnamefont {Schwartz}}, \bibinfo {author} {\bibfnamefont {M.~A.}\ \bibnamefont {Ijaz}}, \bibinfo {author} {\bibfnamefont {X.}~\bibnamefont {Xu}}, \bibinfo {author} {\bibfnamefont {S.}~\bibnamefont {Campbell}}, \bibinfo {author} {\bibfnamefont {G.~T.}\ \bibnamefont {Landi}},\ and\ \bibinfo {author} {\bibfnamefont {M.~S.}\ \bibnamefont {Blok}},\ }\href {https://doi.org/10.48550/arXiv.2512.05237} {{\bibinfo {title} {Analog quantum simulation of the {Lipkin}-{Meshkov}-{Glick} model in a transmon qudit}}} (\bibinfo {year} {2025}),\ \bibinfo {note} {arXiv:2512.05237 [quant-ph]}\BibitemShut {NoStop}%
  \bibitem [{\citenamefont {Makhalov}\ \emph {et~al.}(2019)\citenamefont {Makhalov}, \citenamefont {Satoor}, \citenamefont {Evrard}, \citenamefont {Chalopin}, \citenamefont {Lopes},\ and\ \citenamefont {Nascimbene}}]{makhalov_probing_2019}%
  \BibitemOpen
  \bibfield  {author} {\bibinfo {author} {\bibfnamefont {V.}~\bibnamefont {Makhalov}}, \bibinfo {author} {\bibfnamefont {T.}~\bibnamefont {Satoor}}, \bibinfo {author} {\bibfnamefont {A.}~\bibnamefont {Evrard}}, \bibinfo {author} {\bibfnamefont {T.}~\bibnamefont {Chalopin}}, \bibinfo {author} {\bibfnamefont {R.}~\bibnamefont {Lopes}},\ and\ \bibinfo {author} {\bibfnamefont {S.}~\bibnamefont {Nascimbene}},\ }\bibfield  {title} {{\bibinfo {title} {Probing {Quantum} {Criticality} and {Symmetry} {Breaking} at the {Microscopic} {Level}}},\ }\href {https://doi.org/10.1103/PhysRevLett.123.120601} {\bibfield  {journal} {\bibinfo  {journal} {Physical Review Letters}\ }\textbf {\bibinfo {volume} {123}},\ \bibinfo {pages} {120601} (\bibinfo {year} {2019})}\BibitemShut {NoStop}%
\bibitem [{\citenamefont {Morrison}\ and\ \citenamefont {Parkins}(2008)}]{morrison_dynamical_2008}%
  \BibitemOpen
  \bibfield  {author} {\bibinfo {author} {\bibfnamefont {S.}~\bibnamefont {Morrison}}\ and\ \bibinfo {author} {\bibfnamefont {A.~S.}\ \bibnamefont {Parkins}},\ }\bibfield  {title} {{\bibinfo {title} {Dynamical {Quantum} {Phase} {Transitions} in the {Dissipative} {Lipkin}-{Meshkov}-{Glick} {Model} with {Proposed} {Realization} in {Optical} {Cavity} {QED}}},\ }\href {https://doi.org/10.1103/PhysRevLett.100.040403} {\bibfield  {journal} {\bibinfo  {journal} {Physical Review Letters}\ }\textbf {\bibinfo {volume} {100}},\ \bibinfo {pages} {040403} (\bibinfo {year} {2008})}\BibitemShut {NoStop}%
\bibitem [{\citenamefont {Xu}\ \emph {et~al.}(2020)\citenamefont {Xu}, \citenamefont {Sun}, \citenamefont {Liu}, \citenamefont {Zhang}, \citenamefont {Li}, \citenamefont {Dong}, \citenamefont {Ren}, \citenamefont {Zhang}, \citenamefont {Nori}, \citenamefont {Zheng}, \citenamefont {Fan},\ and\ \citenamefont {Wang}}]{xu_probing_2020}%
  \BibitemOpen
  \bibfield  {author} {\bibinfo {author} {\bibfnamefont {K.}~\bibnamefont {Xu}}, \bibinfo {author} {\bibfnamefont {Z.-H.}\ \bibnamefont {Sun}}, \bibinfo {author} {\bibfnamefont {W.}~\bibnamefont {Liu}}, \bibinfo {author} {\bibfnamefont {Y.-R.}\ \bibnamefont {Zhang}}, \bibinfo {author} {\bibfnamefont {H.}~\bibnamefont {Li}}, \bibinfo {author} {\bibfnamefont {H.}~\bibnamefont {Dong}}, \bibinfo {author} {\bibfnamefont {W.}~\bibnamefont {Ren}}, \bibinfo {author} {\bibfnamefont {P.}~\bibnamefont {Zhang}}, \bibinfo {author} {\bibfnamefont {F.}~\bibnamefont {Nori}}, \bibinfo {author} {\bibfnamefont {D.}~\bibnamefont {Zheng}}, \bibinfo {author} {\bibfnamefont {H.}~\bibnamefont {Fan}},\ and\ \bibinfo {author} {\bibfnamefont {H.}~\bibnamefont {Wang}},\ }\bibfield  {title} {\bibinfo {title} {Probing dynamical phase transitions with a superconducting quantum simulator},\ }\href {https://doi.org/10.1126/sciadv.aba4935} {\bibfield  {journal} {\bibinfo  {journal} {Science Advances}\ }\textbf {\bibinfo {volume} {6}},\
  \bibinfo {pages} {eaba4935} (\bibinfo {year} {2020})}\BibitemShut {NoStop}%
  \bibitem [{\citenamefont {Chinnarasu}\ \emph {et~al.}(2025)\citenamefont {Chinnarasu}, \citenamefont {Poole}, \citenamefont {Phuttitarn}, \citenamefont {Noori}, \citenamefont {Graham}, \citenamefont {Coppersmith}, \citenamefont {Balantekin},\ and\ \citenamefont {Saffman}}]{chinnarasu_variational_2025}%
  \BibitemOpen
  \bibfield  {author} {\bibinfo {author} {\bibfnamefont {R.}~\bibnamefont {Chinnarasu}}, \bibinfo {author} {\bibfnamefont {C.}~\bibnamefont {Poole}}, \bibinfo {author} {\bibfnamefont {L.}~\bibnamefont {Phuttitarn}}, \bibinfo {author} {\bibfnamefont {A.}~\bibnamefont {Noori}}, \bibinfo {author} {\bibfnamefont {T.~M.}\ \bibnamefont {Graham}}, \bibinfo {author} {\bibfnamefont {S.~N.}\ \bibnamefont {Coppersmith}}, \bibinfo {author} {\bibfnamefont {A.~B.}\ \bibnamefont {Balantekin}},\ and\ \bibinfo {author} {\bibfnamefont {M.}~\bibnamefont {Saffman}},\ }\bibfield  {title} {{\bibinfo {title} {Variational {Simulation} of the {Lipkin}-{Meshkov}-{Glick} {Model} on a {Neutral} {Atom} {Quantum} {Computer}}},\ }\href {https://doi.org/10.1103/PRXQuantum.6.020350} {\bibfield  {journal} {\bibinfo  {journal} {PRX Quantum}\ }\textbf {\bibinfo {volume} {6}},\ \bibinfo {pages} {020350} (\bibinfo {year} {2025})}\BibitemShut {NoStop}%
\bibitem [{\citenamefont {Harrington}\ \emph {et~al.}(2022)\citenamefont {Harrington}, \citenamefont {Mueller},\ and\ \citenamefont {Murch}}]{harrington_engineered_2022}%
  \BibitemOpen
  \bibfield  {author} {\bibinfo {author} {\bibfnamefont {P.~M.}\ \bibnamefont {Harrington}}, \bibinfo {author} {\bibfnamefont {E.~J.}\ \bibnamefont {Mueller}},\ and\ \bibinfo {author} {\bibfnamefont {K.~W.}\ \bibnamefont {Murch}},\ }\bibfield  {title} {{\bibinfo {title} {Engineered dissipation for quantum information science}},\ }\href {https://doi.org/10.1038/s42254-022-00494-8} {\bibfield  {journal} {\bibinfo  {journal} {Nature Reviews Physics}\ }\textbf {\bibinfo {volume} {4}},\ \bibinfo {pages} {660} (\bibinfo {year} {2022})}\BibitemShut {NoStop}%
\end{thebibliography}
\end{document}